\documentclass[prl,twocolumn,superscriptaddress,showpacs]{revtex4-1}

\usepackage{amsmath}    
\usepackage{amssymb}    
\usepackage{graphicx}   
\usepackage{verbatim}   
\usepackage{color}      
\usepackage{subfigure}  
\usepackage{hyperref}   
\usepackage{chemfig} 
\usepackage{dsfont}
\usepackage{epsfig}
\begin{document}

\title{Single Electron Dynamics of an Atomic Silicon Quantum Dot \\ on the H-Si$\mathbf{(100)~2\times1}$ Surface}

\author{Marco Taucer}
\affiliation{Department of Physics, University of Alberta, Edmonton, Alberta, Canada, T6G 2G7}
\affiliation{Quantum Silicon Inc, Edmonton, Alberta, Canada, T6G 2M9}

\author{Lucian Livadaru}
\affiliation{Quantum Silicon Inc, Edmonton, Alberta, Canada, T6G 2M9}
\affiliation{National Institute for Nanotechnology, National Research Council of Canada, Edmonton, Alberta, Canada, T6G 2M9}

\author{Paul G. Piva}
\affiliation{Quantum Silicon Inc, Edmonton, Alberta, Canada, T6G 2M9}
\affiliation{National Institute for Nanotechnology, National Research Council of Canada, Edmonton, Alberta, Canada, T6G 2M9}

\author{Roshan Achal}
\affiliation{Department of Physics, University of Alberta, Edmonton, Alberta, Canada, T6G 2G7}
\affiliation{National Institute for Nanotechnology, National Research Council of Canada, Edmonton, Alberta, Canada, T6G 2M9}

\author{Hatem Labidi}
\affiliation{Department of Physics, University of Alberta, Edmonton, Alberta, Canada, T6G 2G7}
\affiliation{National Institute for Nanotechnology, National Research Council of Canada, Edmonton, Alberta, Canada, T6G 2M9}

\author{Jason L. Pitters}
\affiliation{National Institute for Nanotechnology, National Research Council of Canada, Edmonton, Alberta, Canada, T6G 2M9}

\author{Robert A. Wolkow}
\affiliation{Department of Physics, University of Alberta, Edmonton, Alberta, Canada, T6G 2G7}
\affiliation{Quantum Silicon Inc, Edmonton, Alberta, Canada, T6G 2M9}
\affiliation{National Institute for Nanotechnology, National Research Council of Canada, Edmonton, Alberta, Canada, T6G 2M9}

\begin{abstract}
Here we report the direct observation of single electron charging of a single atomic Dangling Bond (DB) on the H-Si(100)~$2 \times 1$ surface. The tip of a scanning tunneling microscope is placed adjacent to the DB to serve as a single electron sensitive charge-detector. Three distinct charge states of the dangling bond, positive, neutral, and negative, are discerned. Charge state probabilities are extracted from the data, and analysis of current traces reveals the characteristic single electron charging dynamics. Filling rates are found to decay exponentially with increasing tip-DB separation, but are not a function of sample bias, while emptying rates show a very weak dependence on tip position, but a strong dependence on sample bias, consistent with the notion of an atomic quantum dot tunnel coupled to the tip on one side and the bulk silicon on the other.
\end{abstract}


\maketitle

\section{Introduction}
Quantum dots are a building block for a range of candidate technologies.
Single electrons can be trapped and manipulated in single- or multiple-quantum dot structures which allows control over occupation down to single electrons\cite{Ciorga2000,Simmons2007}, single electron charge detection\cite{Field1993,Gaudreau2006}, and coherent control of both spatial wavefunctions\cite{Petersson2010,Korkusinski2007} and spin states\cite{Gaudreau2011,Busl2013,Ribeiro2013}. Schemes for employing quantum dot systems have been developed to the level of architectures for both classical\cite{Lent1993,Lu2013} and quantum\cite{Loss1998} information applications. A drawback of most quantum dot systems is the need for cryogenic temperatures, a straightforward consequence of the relatively small charging energies of the quantum dots. As quantum dots are miniaturized, charging energies are increased. Ultimately miniaturized quantum dots are embodied in atomic impurities and atom-scale defects, and these are giving birth to a new arena for technological progress. Recent work on embedded impurities in silicon has delivered impressive single electron devices, demonstrating a single atom Single Electron Transistor (SET)\cite{Fuechsle2012}, coherent spin control\cite{Morello2010}, and optical addressing of single atoms\cite{Yin2013,Smakman2013}. Currently, embedded impurities cannot be placed with atom-scale precision, which is a fundamental limitation for some applications\cite{Kane2005}. By contrast, Dangling Bonds (DBs) on the silicon surface can be fabricated with truly atomic precision\cite{Haider2009,Pitters2011a}, and are thus an attractive candidate for atomic quantum dots. 

On the H-Si(100)~$2 \times 1$ surface, DBs are isolated $\text{sp}^3$ orbitals that do not participate in chemical bonding, and introduce within the bandgap a single surface state. Single DBs exist where there is a single hydrogen atom missing from an otherwise hydrogen-terminated surface.
Localization of charge, variable occupation, strong Coulomb interaction, and the possibility of creating tailor-made structures including tunnel coupling between DBs, leads to a description of DBs as Atomic Silicon Quantum Dots (ASiQDs). There is a body of theoretical work exploring the possibility of using DBs as building blocks for transport and logic devices \cite{[][{ (accepted).}]Yazdi2014,Robles2012,Kawai2012,Livadaru2010}. The potential of using DBs to create functional device elements with tailor-made electronic properties is only just being explored and understood\cite{Bellec2013,Schofield2013,Pitters2011a}, and likewise fabrication is now being optimized (and commercialized)\cite{Goh2011,Chen2012,Kolmer2013}. The discussion that follows will show that the single-electron effects and single-electron charge detection that have become commonplace in many quantum dot systems also exist and can be observed in single DBs. 

\begin{figure}[t]
\setlength{\abovecaptionskip}{2pt}
\setlength{\belowcaptionskip}{-10pt}
\centering
\scalebox{0.30}{
\includegraphics[width=1\textwidth]{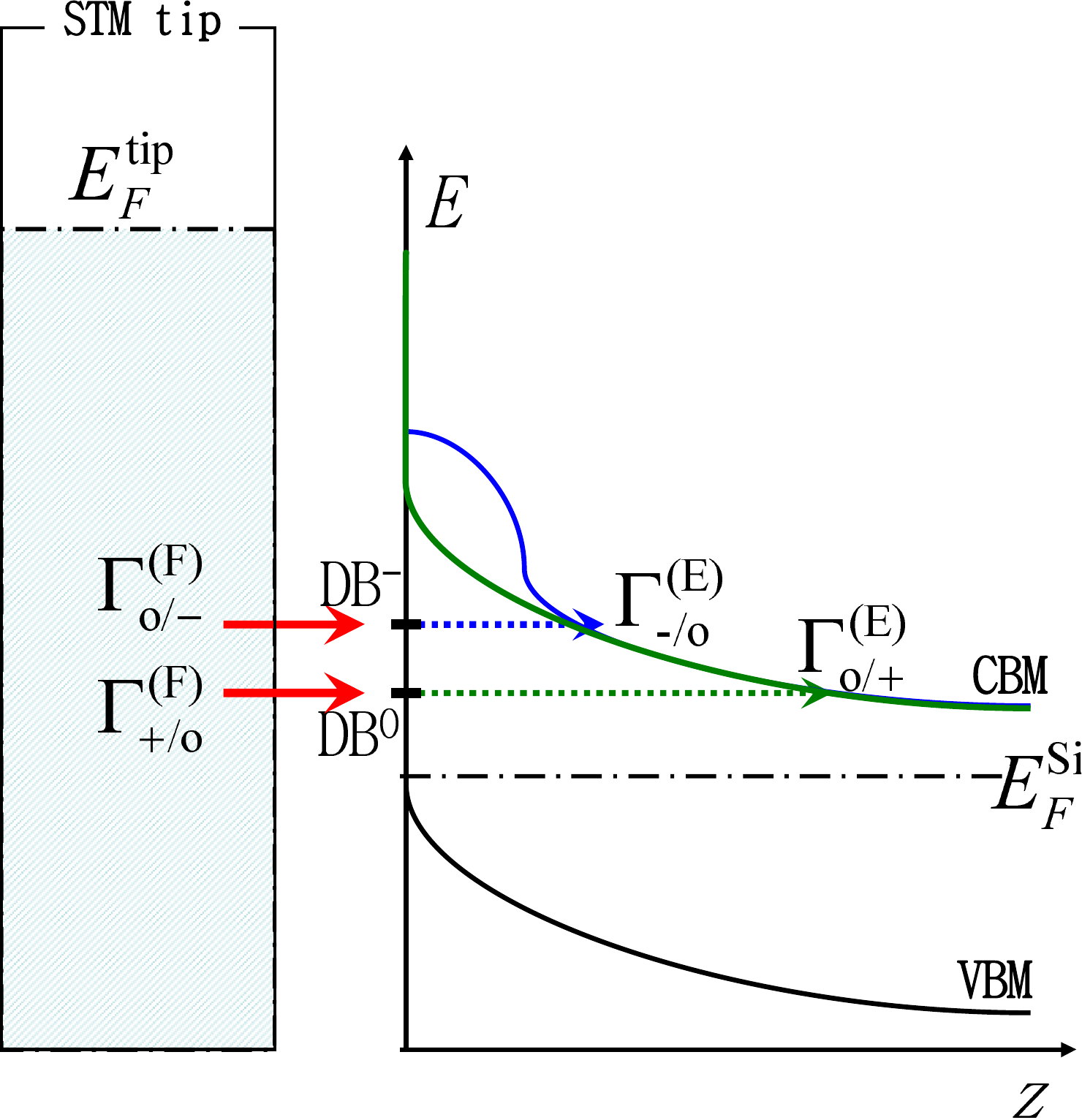}
}
\caption{(Color online) Band diagram representing the non-equilibrium dynamics of STM imaging of a DB. $E_{\rm{F}}^{\rm{tip}}$ and $E_{\rm{F}}^{\rm{Si}}$ label the tip and sample chemical potentials, respectively, and $\Gamma$ labels the filling and emptying processes as described in the text. Tunneling (from tip to DB and from DB to resonant CB levels), dominates both filling and emptying at low temperature, and the DB level moves upward in energy when it is doubly occupied. TIBB and DB-induced band bending (blue curve) both distort the silicon band structure at the surface. The area between the blue curve and the blue arrow indicates the barrier for tunneling from the $\rm{DB}^{-}$ level to resonant CB levels, while the area between the green curve and arrow indicates the tunneling barrier from the $\rm{DB}^{\rm{o}}$ level.}
\label{fig:diagram}
\end{figure}

Generally, the electronic levels of silicon DBs depend on the surface reconstruction and on
occupation; on H-Si(100), the neutral state, $\rm{DB}^{\rm{o}}$, is estimated at 0.35~eV and the negative state, $\rm{DB}^{-}$, is estimated at 0.85~eV above the valence band\cite{Livadaru2010} (Figure \ref{fig:diagram}). Although there are three distinct charge states, we only refer to two energy levels, the so-called neutral and negative energy levels. These are the Fermi energy thresholds above which the DB becomes neutral and negative respectively. 

Despite the central role of single-electron dynamics in STM imaging of DBs\cite{Livadaru2011} as well as in potential DB-based atom-scale devices\cite{Kawai2012}, until now, they have not been directly observed in an STM experiment. Here, we report direct observation of single-electron charging dynamics of DBs. The dynamics are consistent with a model of non-equilibrium charging, in which the DB acts as the island of an SET, tunnel-coupled to the STM tip and to the silicon bulk. The tip-sample tunnel junction is also gated by the charge of the nearby single DB, so that the total tunneling current acts as a single-electron sensitive charge sensor, reminiscent of the charge detection scheme commonly used in more familiar quantum dot systems, such as those formed in two-dimensional electron gas (2DEG) systems. Besides being of fundamental relevance for the quantum dynamics of single electrons in ultimately small quantum dot junctions, our study provides insight into the inherent characteristics and limits of molecular-scale devices, including charge qubits, SETs, and other examples of quantum engineering\cite{Schofield2013} on semiconductor surfaces in realistic conditions, i.e. presence of control electrodes and bulk continua.

For these experiments, we used an Omicron LT-STM operated at 4.2~K. The tungsten tip was prepared by electrochemical etching followed by electron beam heating and field ion microscopy cleaning and sharpening\cite{Rezeq2006}. The sample was cleaved from a 3-4~$\rm{m}\Omega \cdot \rm{cm}$ $n$-type As-doped Si(100) wafer, and was cleaned by heating several times to roughly $1250^{\rm{o}}\rm{C}$, and H-terminated at $330^{\rm{o}}\rm{C}$\cite{Boland1993}. 
The high temperatures used to clean the sample are known to deplete the dopants near the surface\cite{Pitters2012}.

\begin{figure}[b]
\setlength{\abovecaptionskip}{3pt}
\setlength{\belowcaptionskip}{0pt}
\centering
\scalebox{0.45}{
\includegraphics[width=1\textwidth]{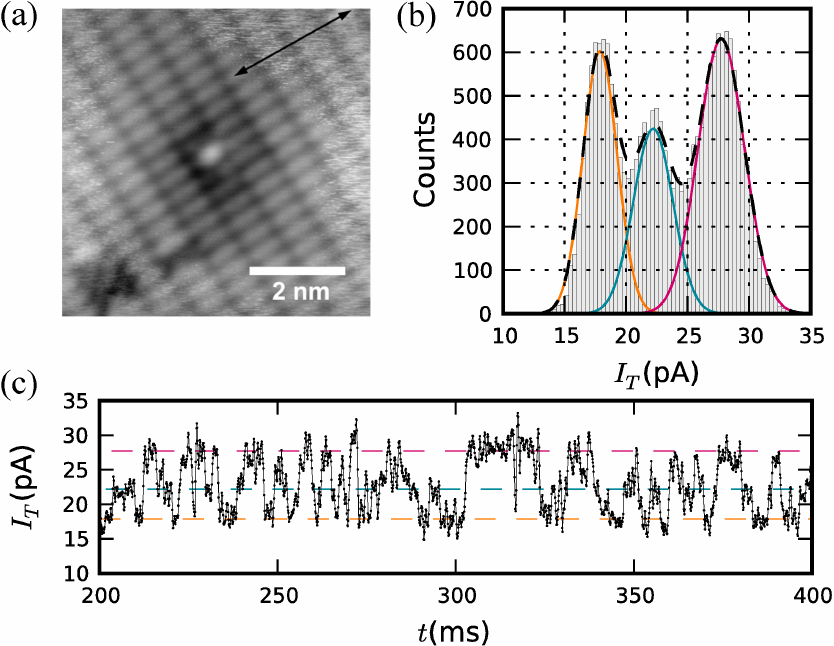}
}
\caption{(Color online) \textbf{(a)} Topographical STM image of a single DB taken with $V_S=1.4~\rm{V}$ and $I_T=20~\rm{pA}$. The double-ended arrow shows the range of lateral positions used to acquire the data shown in Figure \ref{fig:cmaps}. \textbf{(b)} Histogram of current measurements with the tip at a constant height and a constant voltage of $V_S=1.45~\rm{V}$ positioned 3.14nm from the DB. The peak at lowest current corresponds to the negative charge state, while the peaks at intermediate and highest current correspond to the neutral and positive charge states. \textbf{(c)} An example of a current-time trace. The sampling rate is 10~kHz and the entire trace (not shown) is 2s in length. }
\label{fig:noise}
\end{figure}

The dark ``halos" which surround DBs imaged in unoccupied states (see Figure \ref{fig:noise}(a) and Figure \ref{fig:RT} have been attributed to upward band bending near a negatively charged DB. Yet certain features of these halos --- namely, the sharpness of the edge of the halo and the dependence of halo shape on tip geometry and voltage --- require detailed consideration. A theory of STM of DBs which captures these features was put forth by Livadaru \textit{et al.}\cite{Livadaru2011}. In unoccupied state imaging, Tip-Induced Band Bending (TIBB) tends to empty nearby states (including the DB). But at the same time, electrons tunnel from the tip to the unoccupied sample energy levels. When an electron is injected into a localized state, the dynamics which would re-establish thermal equilibrium in the sample can be relatively slow, and the equilibrium picture of STM no longer applies. According to this non-equilibrium picture of STM imaging of DBs, the charge state of the DB is determined by the competition of filling and emptying processes, as shown in Figure \ref{fig:diagram}.

\begin{figure*}[t!]
\setlength{\abovecaptionskip}{-6pt}
\setlength{\belowcaptionskip}{-10pt}
\centering
\scalebox{0.9}{
\includegraphics[width=1\textwidth]{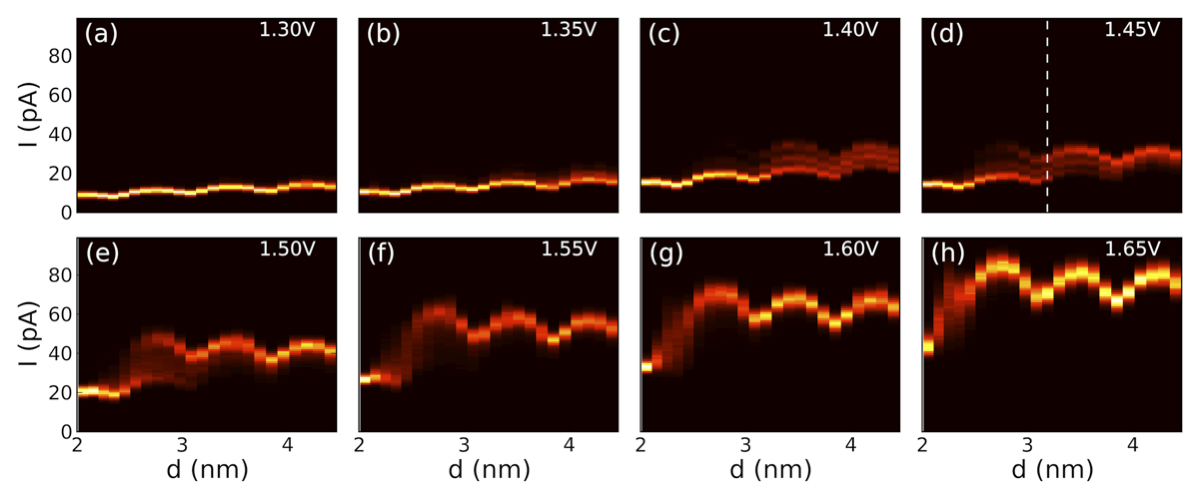}}
\caption{(Color online) Colormaps showing frequency of current measurements as a function of tip-DB separation and current from (a)~1.30~V to (h)~1.65~V. Colormap intensity is proportional to the number of instances of a particular measurement of current at a particular position. Any vertical slice of any colormap gives a histogram whose total integral is $2~\rm{s} \times 10~\rm{kHz} = 20 000~\rm{samples}$. In particular, the dotted line in (d) corresponds to the data shown in Fig. \ref{fig:noise}(b).}
\label{fig:cmaps}
\end{figure*}

At low temperature, many thermal processes become negligible, so that dynamics can be slow enough to be within the STM pre-amplifier bandwidth ($\sim$10~kHz). Figure \ref{fig:noise}(a) shows a topographical unoccupied state image of a DB at 4.2~K. At these conditions, the edge of the halo is no longer sharp, but instead shows a distinctive streaky noise. When the tip is positioned in the halo region, and the tip height is held constant, the measurement of current as a function of time shows unusual jumps to discrete values, as seen in Figure \ref{fig:noise}(c). Such current steps are absent when the tip is far from from any DBs. The histogram of current measurements shown in Figure \ref{fig:noise}(b) demonstrates that there are precisely three dominant current values. We identify these as corresponding to the negative (doubly occupied), neutral (singly occupied), and positive (unoccupied) charge states of the DB. Each charge state of the DB causes a different DB-induced band bending under the tip apex, and thereby creates a different current from tip to sample.

The electron dynamics represented in Figure \ref{fig:noise}(b) and (c) are for a particular tip position and voltage, but in general the dynamics and populations of the three charge states will depend on these parameters. Each panel in Figure \ref{fig:cmaps} compactly shows a collection of histograms at different lateral positions crossing the edge of a DB halo, for a particular tip voltage. Colormap intensity is proportional to the number of counts at a particular current and tip-DB separation, and tip height is constant for all the data presented. At all voltages there is a periodic modulation in current as a function of position, due to the surface topography; as the tip moves laterally at constant height, the tip-sample distance is modulated because of the periodicity of the silicon surface. The striking feature is the appearance of current instability reflected by the broadening and/or existence of multiple current levels at particular voltages and positions.


At the lowest sample voltage (Figure \ref{fig:cmaps}(a) at 1.30~V), the DB is in a single charge state at all tip positions. The STM current decreases as the tip moves toward the DB (aside from the abovementioned periodic modulation due to topography), indicating upward band bending near the DB, consistent with a negative charge state. As the tip bias is increased (Figure \ref{fig:cmaps}(b-h)), two additional charge states become visible, which we identify as the neutral and positive DB states. For voltages greater than 1.35~V, there is a transition region in which all three charge states are visible, with the positive DB charge state becoming dominant at larger tip-DB separations. Above 1.50~V, transitions occur on a timescale which competes with the data acquisition rate, so that the three states become blurred and eventually averaged. At 1.55~V and above, the high-current peak dominates for most tip positions, and here we see that current \emph{increases} as the tip moves toward the DB, indicating \emph{downward} band bending near the DB, consistent with a positive charge state. All traces show a low-current value at the smallest tip-DB separation because direct tunneling from the tip to the DB becomes dominant, in turn causing negative charging of the DB.

Looking at Figure \ref{fig:noise}(c) we can see that the $I(t)$ trace contains more information than just the probability of the various charge states as a function of bias and tip position. There is dynamical information. The trace consists of plateaux of various lengths at each of the three currents corresponding to the three possible charge states. Dynamical information can in principle be extracted by measuring the lengths of these plateaux as well as which states they transition to. This would give the transition rates, $\Gamma_{-/\text{o}}^{(\rm{E})}$, $\Gamma_{\text{o}/+}^{(\rm{E})}$, $\Gamma_{\text{o}/-}^{(\rm{F})}$, and $\Gamma_{+/\text{o}}^{(\rm{F})}$, for the kinetic scheme
\begin{equation}
\schemestart
$\rm{DB}^{-}$
\arrow(A--B){<=>[$\Gamma_{-/\text{o}}^{(\rm{E})}$][$\Gamma_{\text{o}/-}^{(\rm{F})}$]}
$\rm{DB}^{o}$
\arrow(A--B){<=>[$\Gamma_{\text{o}/+}^{(\rm{E})}$][$\Gamma_{+/\text{o}}^{(\rm{F})}$]}
$\rm{DB}^{+}$ \quad,
\schemestop
\label{eqn:scheme}
\end{equation}
which assumes that there is no direct transition between the negative state and positive states. This assumption amounts to neglecting any particular two-electron filling or emptying processes, but does not preclude the possibility of going very rapidly from the positive to the negative charge state (or vice versa) via two single-electron processes in quick succession. The superscripts $(\rm{F})$ and $(\rm{E})$ indicate filling and emptying rates. However, determining these rates by simply measuring the lengths of plateaux in Figure \ref{fig:noise}(c) turns out to be problematic, since the noise in the plateaux is comparable with their separation. Motivated by this, we take an approach developed by Hoffmann and Woodside\cite{Hoffmann2013} called signal-pair analysis, which is based on calculating the evolution of probability distributions in time, and thereby extracting the transition rates between states even if their signals overlap significantly. This analysis combines earlier work on single-molecule fluorescence studies\cite{Hoffmann2011}, and a signal-pair correlation approach to analyzing structural dynamics of proteins\cite{Hoffmann2011a}. The procedure is explained in the supplemental materials section.

\begin{figure}[h]
\setlength{\abovecaptionskip}{-5pt}
\setlength{\belowcaptionskip}{-5pt}
\centering
\scalebox{0.45}{\includegraphics[width=1\textwidth]{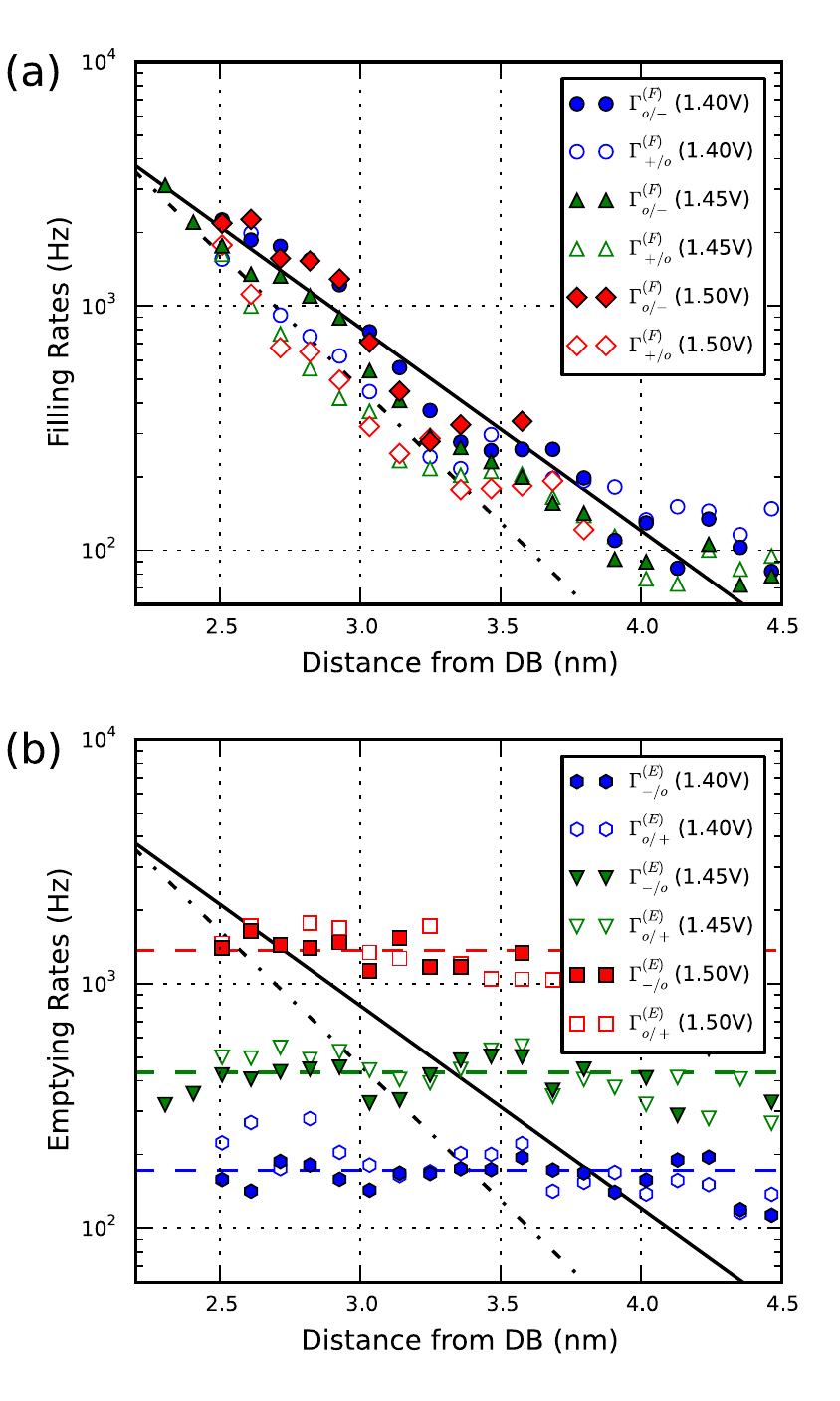}}
\caption{(Color online) 
\textbf{(a)} Experimentally measured filling rates as a function of lateral tip distance from DB for three different voltages. The dashed dotted line indicates the exponential fit to the filling rate of the neutral DB energy level, while the solid line indicates the fit for the negative DB level.
\textbf{(b)} Experimentally measured emptying rates as a function of lateral tip distance from DB for three different voltages. For each voltage, emptying rates have a weak dependence on tip position. The dashed coloured lines show the average emptying rate for each voltage, while the dashed-dotted and solid black lines show the same fits to the filling rates as shown in (a).}
\label{fig:rates}
\end{figure}

The extracted filling and emptying rates are shown in Figure \ref{fig:rates}.  Figure \ref{fig:rates}(a) shows the dependence of filling rates on tip position for three different voltages. There is an exponential decay in the filling rate with increasing tip-DB separation, with values from roughly 3kHz to 50Hz, with no clear systematic dependence of filling rates on voltage. This is consistent with the prediction that direct tunneling from tip to DB dominates the filling of the DB, assuming a constant density of tip states over the energy range of interest. The exponential fits to the filling rates of the neutral and negative charge states are shown as a black dashed-dotted line and a black solid line, respectively. Their decay rates are $k_{\rm{o/-}}^{(\rm{F})} = 1.91~\rm{nm}^{-1}$ and $k_{\rm{+/o}}^{(\rm{F})} = 2.54~\rm{nm}^{-1}$. We attribute the slower decay of the negative charge state to the upward shift of the $\rm{DB}^{-}$ energy level with respect to that of $\rm{DB}^{\rm{o}}$, resulting in a smaller ionization potential. 

Figure \ref{fig:rates}(b) shows emptying rates. In contrast to figure \ref{fig:rates}(a), we see a strong voltage dependence and a very weak position dependence. While calculations of room temperature STM of DBs\cite{Livadaru2011} found thermal emission of electrons from the DB level to dominate emptying, this process is nearly eliminated at 4.2~K. We instead consider the dominant mechanism at low temperature to be tunneling from the DB energy level to distant but resonant CB levels. As the bias is increased, TIBB is also increased, while the associated barrier for an electron on the DB to tunnel to the CB becomes narrower. Specifically, we find relatively flat emptying rates of 172~Hz, 434~Hz, and 1369~Hz for sample voltages of 1.40~V, 1.45~V, and 1.50~V respectively. The average emptying rate for each of the three voltages is shown as a horizontal dashed line. The weak dependence of emptying rates on tip position is an indication that TIBB is relatively uniform on the scale considered here, as expected.

We can now see that the edge of the DB halo is the point at which filling rates overtake emptying rates. This corresponds to the intersection of the horizontal dashed lines (emptying) in figure \ref{fig:rates}(b) with the exponential solid and dashed dotted lines (filling). As voltage is increased, the emptying rate, which is nearly flat with respect to position, increases while the filling rate remains an unchanged exponential. The point of intersection (edge of the halo) thus moves toward the DB. This is consistent with our routine observation of a DB halo size which decreases with increasing bias (not shown).

In conclusion, we have shown that single-electron dynamics are directly observable in STM of single DBs when the tunnel junction between the tip and the sample acts as a single-electron sensitive charge detector. We directly resolve the three possible charge states, negative, neutral, and positive, of a DB. The dynamics extracted from current traces are consistent with a non-equilibrium model in which the DB acts as an atomic quantum dot, tunnel-coupled both to the tip and to the bulk silicon, and occupation is determined by the competition of filling from the tip and emptying to the bulk.  These results show that the charge state of an atomic quantum dot can be manipulated and read by nearby electrodes. There is no fundamental reason why the single atom charge state sensing demonstrated here cannot in future be implemented in an STM-free, lithographic structure.

\section{Supplemental Material}
\subsection{Data Analysis: Extracting Transition Rates}

Given a dataset like the one shown in Figures 2(b) and (c), it is possible to determine the total fraction of time spent in the negative, neutral, or positive charge state --- that is, we can determine the probabilities $P_{-}$, $P_{\rm{o}}$, and $P_{+}$, whose sum must of course be one, for the DB to be found in each charge state at an arbitrary time. We can represent these three probabilities more concisely as a probability ``vector", $\mathbf{P}\equiv \left(P_{-},P_{\rm{o}},P_{+}\right)^{\mathsf{T}}$.
The integral area of each of the three Gaussian fits shown in figure 2(b) of the manuscript is proportional to the probability for that charge state, which we refer to as the ``steady state" probability, whose vector is $\mathbf{P}_{\rm{ss}}$.

The widths of the Gaussians that make up the histograms are determined by the noise in tunneling current (in the absence of transitions between charge states), which is dominated by two contributions: the intrinsic pre-amplifier noise, and tip-height noise. The intrinsic noise of the STM pre-amplifier, $\delta I_{\rm{pre}}$, is a constant. The noise in tip height, $\delta z$, is also a constant, but its contribution to the current noise, $\delta I_z$, is given by $ \delta I_z \approx \left| \frac{dI}{dz} \right| \delta z $.
But since the current is an exponential function of tip height, we have $ \left| \frac{dI}{dz} \right| \propto I$, which leads to the conclusion that the noise in tip height contributes a noise in current that is proportional to the mean current: $ \delta I_z \propto \overline{I} $. Taking these to be the dominant contributors to current noise, we are led to the conclusion that Gaussian width should be well approximated by a linear function of mean. By constraining the Gaussian widths, the number of free parameters in our triple Gaussian fits was decreased, which allowed us to analyze a wider range of data (\textit{e.g.} cases in which some peaks would otherwise be hard to resolve). In the data presented here, good fits were found by constraining the widths to be the following linear function of the mean:
\begin {equation}
\sigma = 0.5~\rm{pA} + 0.05\mu . \nonumber
\end{equation}

If we denote the dataset corresponding to a particular current trace (at a particular position and voltage) as $\mathcal{D} \equiv \{ I(t)\}$, then $\mathbf{P}_{\rm{ss}}$ is obtained from the triple Gaussian fit to the entire dataset, $\mathcal{D}$. We may also define a probability vector for any subset of this dataset, $\mathcal{S} \subseteq \mathcal{D}$, by fitting the histogram corresponding to $\mathcal{S}$ using a constrained triple Gaussian fit whose widths \textit{and} means are fixed by the fit to $\mathcal{D}$, but whose amplitudes are free parameters.

\begin{figure}[h]
\centering
\scalebox{0.47}{\includegraphics[width=1\textwidth]{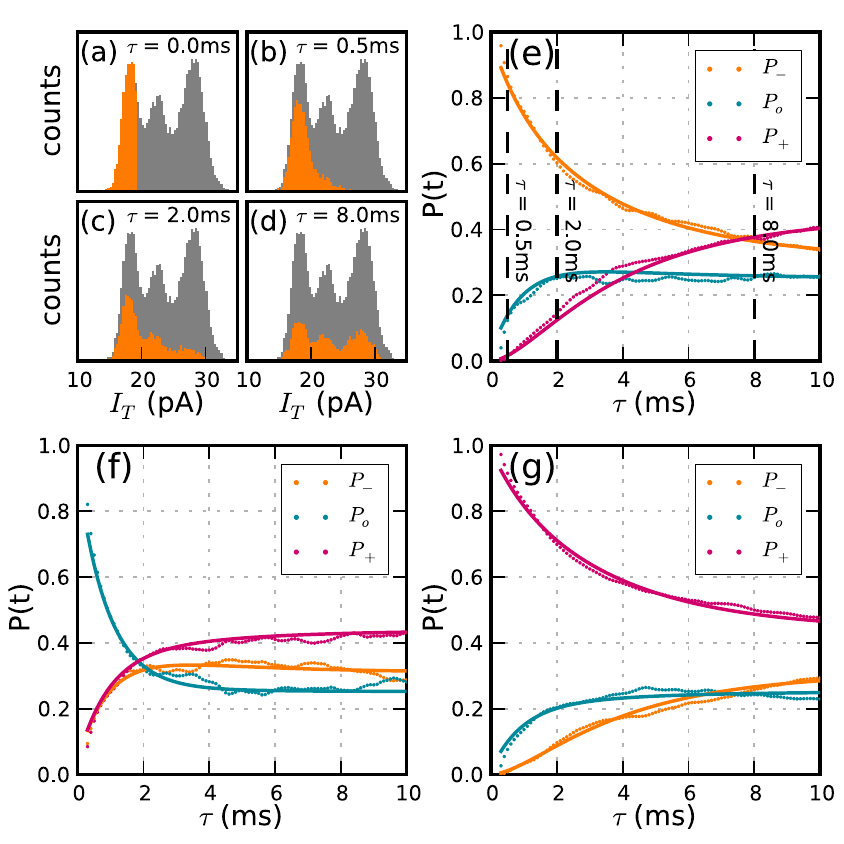}}
\caption{\textbf{(a-d)} The grey histogram is the histogram corresponding to the data set, $\mathcal{D}$, shown in Figures 2(b) and (c). The superimposed orange histogram in (a) shows the subset, $\mathcal{S}(\tau=0)$, chosen such that $I(t) \leq 19.0~\rm{pA}$. The superimposed histograms in (b-d) show the subsets $\mathcal{S}(\tau=0.5~\rm{ms})$, $\mathcal{S}(\tau=2.0~\rm{ms})$, and $\mathcal{S}(\tau=8.0~\rm{ms})$ respectively. \textbf{(e)} The three components of $\mathbf{P}(\tau)$ are plotted as a function of $\tau$ for the subset shown in (a), that is $\mathcal{S}(0) \equiv \{ I(t) :  \mathnormal{I(t)} \leq 19.0~\rm{pA} \}$. The vertical dashed lines show the components of $\mathbf{P}$ corresponding to the subsets shown in (b-d). \textbf{(f)} Likewise, the three components of $\mathbf{P}(\tau)$ for a different initial subset $\mathcal{S}(0) \equiv \{ I(t) : 20.6~\rm{pA} \leq \mathnormal{I(t)} \leq 23.8~\rm{pA} \}$. \textbf{(g)} The three components of $\mathbf{P}(\tau)$ for $\mathcal{S}(0) \equiv \{ I(t) : 25.4~\rm{pA} \leq \mathnormal{I(t)} \}$.}
\label{fig:corr}
\end{figure}

The grey histograms shown in Figures \ref{fig:corr}(a-d) are all exactly the histogram shown in Figure 2(b). We denote the corresponding data set $\mathcal{D}$. In Figure \ref{fig:corr}(a), the orange part of the histogram shows the distribution corresponding to the subset of data points for which current is less than 19pA, that is $\mathcal{S}(0) \equiv \{ I(t) : \mathnormal{I(t)} \leq 19~\rm{pA} \}$. We can also consider the distribution of points which occur exactly a time $\tau$ later than the original subset, that is $\mathcal{S}(\tau) \equiv \{ I(t+\tau) : \mathnormal{I(t)} \leq 19~\rm{pA} \}$. The orange histograms in Figures \ref{fig:corr}(b-d) show the distributions of $\mathcal{S}(\tau)$ for $\tau$ equal to 0.5~ms, 2.0~ms, and 8.0~ms respectively. As $\tau$ increases, we see that the distribution first spreads to take roughly the shape of the negative charge state Gaussian, and subsequently the amplitude of that peak decreases as the other two increase. At 8.0ms the orange distribution, $\mathcal{S}(\tau=8.0~\rm{ms})$ is approaching the steady state, which is to say that it becomes a scaled down version of the grey one, $\mathcal{D}$.

There is a probability vector, $\mathbf{P}(\tau)$, corresponding to each subset, $\mathcal{S}(\tau)$, which is determined by the constrained triple Gaussian fit described above. The subset shown in Figure \ref{fig:corr}(a) was chosen so that $P_{-}(0) \approx 1$, which is to say that the constrained Gaussian fit to $\mathcal{S}(\tau \simeq 0)$ gives a considerable amplitude to the negative peak with almost none for the neutral and positive peaks. The evolution of the probability vector for this subset is shown in figure \ref{fig:corr}(e). This shows, again, that the negative charge state probability drops from 1 as the other two probabilities increase, and all three tend towards their steady state values. Figures \ref{fig:corr}(f) and (g) show the evolution of $\mathbf{P} (\tau)$ for subsets corresponding to the neutral and positive charge states.

The theoretical prediction for $\mathbf{P}(\tau)$ is based on a set of coupled differential equations, corresponding to the kinetic scheme
\begin{equation}
\schemestart
$\rm{DB}^{-}$
\arrow(A--B){<=>[$\Gamma_{-/\text{o}}^{(\rm{E})}$][$\Gamma_{\text{o}/-}^{(\rm{F})}$]}
$\rm{DB}^{o}$
\arrow(A--B){<=>[$\Gamma_{\text{o}/+}^{(\rm{E})}$][$\Gamma_{+/\text{o}}^{(\rm{F})}$]}
$\rm{DB}^{+}$ \quad,
\schemestop
\label{eqn:scheme}
\end{equation}
 which can be concisely expressed as
\begin{eqnarray}
&\frac{d}{d\tau}\mathbf{P}(\tau)=\mathds{M}\mathbf{P}(\tau) \quad ; \\
\nonumber \\
&\mathds{M}\equiv 
\left( \begin{array}{ccc}
-\Gamma_{-/\text{o}}^{(\rm{E})} & \Gamma_{\text{o}/-}^{(\rm{F})} & 0 \\
\Gamma_{-/\text{o}}^{(\rm{E})} & -\Gamma_{\text{o}/+}^{(\rm{E})}-\Gamma_{\text{o}/-}^{(\rm{F})} & \Gamma_{+/\text{o}}^{(\rm{F})} \\
0 & \Gamma_{\text{o}/+}^{(\rm{E})} & -\Gamma_{+/\text{o}}^{(\rm{F})} \end{array} \right).
\nonumber
\label{eqn:DEcoupled}
\end{eqnarray}
It follows from this relation that the filling and emptying rates are constrained by the steady state probabilities through the two relations
\begin{equation}
\frac{ \Gamma_{-/\text{o}}^{(\rm{E})} }{ \Gamma_{\text{o}/-}^{(\rm{F})} } = \frac{ P_{\text{o}}^{\rm{(ss)}} }{ P_{-}^{\rm{(ss)}} } 
\qquad \text{and} \qquad
\frac{ \Gamma_{\text{o}/+}^{(\rm{E})} }{ \Gamma_{+/\text{o}}^{(\rm{F})} } = \frac{ P_{+}^{\rm{(ss)}} }{ P_{\text{o}}^{\rm{(ss)}} }.
\label{eqn:rateconstants}
\end{equation}
The set of coupled differential equations is uncoupled by diagonalizing the matrix $\mathds{M}$ --- that is, finding a diagonal matrix, $\mathds{J}$, such that for some transformation matrix, $\mathds{S}$, we have $\mathds{M}=\mathds{S}\mathds{J}\mathds{S}^{-1}$. This gives the time evolution of the probability vector as
\begin{equation}
\mathbf{P}(\tau)=\mathds{S}e^{\mathds{J}\tau}\mathds{S}^{-1}\mathbf{P}(0).
\label{eqn:P(tau)}
\end{equation}
Any matrix of the form of $\mathds{M}$ has at least one eigenvalue equal to zero with the other two less than or equal to zero. This means that each component of $\mathbf{P}(\tau)$ is comprised of a constant term (the steady state probability for that charge state) plus two decaying exponentials. The solid curves in Figures \ref{fig:corr}(e-g) are fits to the data using equation \ref{eqn:P(tau)}, constrained by the relations \ref{eqn:rateconstants}. All nine curves are fit using only two free parameters. These fits give the four rates for the transitions between states.
 
This analysis was repeated for all current traces where multiple charge states could be resolved. 

\subsection{Room Temperature Results}

\begin{figure}[b]
\centering
\scalebox{0.35}{
\includegraphics[width=1\textwidth]{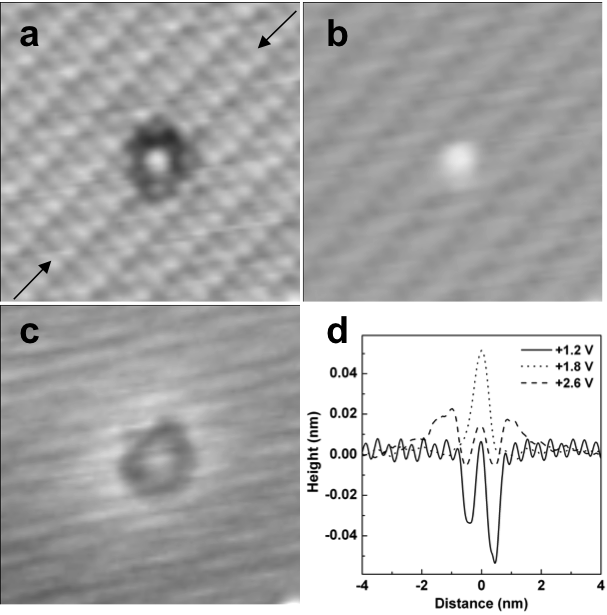}
}
\caption{Constant-current empty-state STM imaging of DBs on n-type H:Si(100) at 300 K. (a) $V_S= +1.2 \rm{V}$. H-silicon within $\sim0.8\rm{nm}$ of the DB (image centre) images with depressed height. (b) $V_S = +1.8 \rm{V}$. DB images as a single bright protrusion. The height of the surrounding H-silicon is unperturbed. (c) $V_S = +2.6 \rm{V}$. DB images as a slight protrusion. H-silicon height within $\sim0.8\rm{nm}$ of the DB is weakly perturbed. Beyond $\sim0.9\rm{nm}$, the H-silicon displays an abrupt increase in imaging height ($\sim0.02 \rm{nm}$) which decays with increasing distance from the DB centre. (d) Topographic cross sections (0.5 nm wide) extracted along the central H-silicon dimer row (indicated by black arrows in (a)) and across the DB centre. Tunnel current: 20 pA. Image areas: $\sim 6 \times 6 \rm{nm}^2$.}
\label{fig:RT}
\end{figure}

The above description of DB halos readily explains systematic variations in the appearance of dangling bonds in empty state imaging routinely observed as a function of bias at room temperature. At low sample biases, as in Figure \ref{fig:RT}(a), the dark halo which surrounds the dangling bond at +1.2 V can be understood as a natural consequence of the DB charge state changing as the tip approaches the DB.  While the tip-DB separation remains greater than ~ 1 nm, TIBB raises the doubly occupied DB level above the Fermi level, leaving the DB charge neutral.  As a result of the lack of local bandbending induced by the DB, the H-silicon images without topographical distortion over this region (see also the +1.2 V topograph in Fig. \ref{fig:RT}(d)).  As the tip-DB separation reaches ~1 nm, however, direct tunneling from the tip causes the DB's negative state to be filled faster than it can empty. The DB therefore takes on a negative charge state, leading to upward local bandbending, and the appearance of a depression or dark halo in the vicinity of the DB.  

As the bias is raised, the tip-sample separation increases to maintain the same tunneling current, decreasing tunneling from tip to DB, while at the same time increasing the field effect of the tip, and associated TIBB.  As a result, the diameter of the dark halo gradually decreases (not shown) until as in Figure \ref{fig:RT}(b), the dark halo disappears altogether.  At this higher bias of + 1.8 V, the decrease in the fraction of current injected from the tip to the DB, as well as the increased emptying rate of the DB (driven by the increased TIBB at the DB), leads to a situation where the DB is on average neutral independent of the tip-DB separation.  The dark halo is now completely absent (see also the + 1.8 V topograph in Figure \ref{fig:RT}(d)). While near the DB there may still be charging dynamics which are faster than the bandwidth of the preamplifier, the time-average of these shows a neutral DB.

When the bias is raised further to + 2.6 V (Figure \ref{fig:RT}(c) and +2.6 V topograph in Figure \ref{fig:RT}(d)), a new and different halo emerges. In this case, the H-Si surface appears to slope upward as the tip approaches the DB, but abruptly drops to a height which is comparable to the height of the unperturbed surface. We can understand this as resulting from a continuation of the trends discussed so far. The increased level of TIBB tends to empty the DB centre, leading to a positive charge state when the tip is at an intermediate distance. This accounts for the brightening (i.e. increased imaging height) of the silicon in the vicinity of the DB. At even smaller tip-DB separations, direct tunneling from the tip to the DB becomes competitive and restores the DB to a neutral state on average, creating the new halo.  Instead of imaging below the plane of the unperturbed H-silicon surface as in Figure \ref{fig:RT}(a), the bottom of the halo (~1 nm from the DB) images with roughly the same height as the unperturbed surface far (³ ~4 nm) from the DB.

\begin{acknowledgments}
We would like to thank Armin Hoffmann and Michael Woodside for very valuable discussions, and Martin Cloutier and Mark Salomons for their technical expertise.
\end{acknowledgments}


\begin{thebibliography}{35}%
\makeatletter
\providecommand \@ifxundefined [1]{%
 \@ifx{#1\undefined}
}%
\providecommand \@ifnum [1]{%
 \ifnum #1\expandafter \@firstoftwo
 \else \expandafter \@secondoftwo
 \fi
}%
\providecommand \@ifx [1]{%
 \ifx #1\expandafter \@firstoftwo
 \else \expandafter \@secondoftwo
 \fi
}%
\providecommand \natexlab [1]{#1}%
\providecommand \enquote  [1]{``#1''}%
\providecommand \bibnamefont  [1]{#1}%
\providecommand \bibfnamefont [1]{#1}%
\providecommand \citenamefont [1]{#1}%
\providecommand \href@noop [0]{\@secondoftwo}%
\providecommand \href [0]{\begingroup \@sanitize@url \@href}%
\providecommand \@href[1]{\@@startlink{#1}\@@href}%
\providecommand \@@href[1]{\endgroup#1\@@endlink}%
\providecommand \@sanitize@url [0]{\catcode `\\12\catcode `\$12\catcode
  `\&12\catcode `\#12\catcode `\^12\catcode `\_12\catcode `\%12\relax}%
\providecommand \@@startlink[1]{}%
\providecommand \@@endlink[0]{}%
\providecommand \url  [0]{\begingroup\@sanitize@url \@url }%
\providecommand \@url [1]{\endgroup\@href {#1}{\urlprefix }}%
\providecommand \urlprefix  [0]{URL }%
\providecommand \Eprint [0]{\href }%
\providecommand \doibase [0]{http://dx.doi.org/}%
\providecommand \selectlanguage [0]{\@gobble}%
\providecommand \bibinfo  [0]{\@secondoftwo}%
\providecommand \bibfield  [0]{\@secondoftwo}%
\providecommand \translation [1]{[#1]}%
\providecommand \BibitemOpen [0]{}%
\providecommand \bibitemStop [0]{}%
\providecommand \bibitemNoStop [0]{.\EOS\space}%
\providecommand \EOS [0]{\spacefactor3000\relax}%
\providecommand \BibitemShut  [1]{\csname bibitem#1\endcsname}%
\let\auto@bib@innerbib\@empty
\bibitem [{\citenamefont {Ciorga}\ \emph {et~al.}(2000)\citenamefont {Ciorga},
  \citenamefont {Sachrajda}, \citenamefont {Hawrylak}, \citenamefont {Gould},
  \citenamefont {Zawadzki}, \citenamefont {Jullian}, \citenamefont {Feng},\
  and\ \citenamefont {Wasilewski}}]{Ciorga2000}%
  \BibitemOpen
  \bibfield  {author} {\bibinfo {author} {\bibfnamefont {M.}~\bibnamefont
  {Ciorga}}, \bibinfo {author} {\bibfnamefont {A.~S.}\ \bibnamefont
  {Sachrajda}}, \bibinfo {author} {\bibfnamefont {P.}~\bibnamefont {Hawrylak}},
  \bibinfo {author} {\bibfnamefont {C.}~\bibnamefont {Gould}}, \bibinfo
  {author} {\bibfnamefont {P.}~\bibnamefont {Zawadzki}}, \bibinfo {author}
  {\bibfnamefont {S.}~\bibnamefont {Jullian}}, \bibinfo {author} {\bibfnamefont
  {Y.}~\bibnamefont {Feng}}, \ and\ \bibinfo {author} {\bibfnamefont
  {Z.}~\bibnamefont {Wasilewski}},\ }\href {\doibase
  10.1103/PhysRevB.61.R16315} {\bibfield  {journal} {\bibinfo  {journal}
  {Physical Review B}\ }\textbf {\bibinfo {volume} {61}},\ \bibinfo {pages}
  {R16315} (\bibinfo {year} {2000})}\BibitemShut {NoStop}%
\bibitem [{\citenamefont {Simmons}\ \emph {et~al.}(2007)\citenamefont
  {Simmons}, \citenamefont {Thalakulam}, \citenamefont {Shaji}, \citenamefont
  {Klein}, \citenamefont {Qin}, \citenamefont {Blick}, \citenamefont {Savage},
  \citenamefont {Lagally}, \citenamefont {Coppersmith},\ and\ \citenamefont
  {Eriksson}}]{Simmons2007}%
  \BibitemOpen
  \bibfield  {author} {\bibinfo {author} {\bibfnamefont {C.~B.}\ \bibnamefont
  {Simmons}}, \bibinfo {author} {\bibfnamefont {M.}~\bibnamefont {Thalakulam}},
  \bibinfo {author} {\bibfnamefont {N.}~\bibnamefont {Shaji}}, \bibinfo
  {author} {\bibfnamefont {L.~J.}\ \bibnamefont {Klein}}, \bibinfo {author}
  {\bibfnamefont {H.}~\bibnamefont {Qin}}, \bibinfo {author} {\bibfnamefont
  {R.~H.}\ \bibnamefont {Blick}}, \bibinfo {author} {\bibfnamefont {D.~E.}\
  \bibnamefont {Savage}}, \bibinfo {author} {\bibfnamefont {M.~G.}\
  \bibnamefont {Lagally}}, \bibinfo {author} {\bibfnamefont {S.~N.}\
  \bibnamefont {Coppersmith}}, \ and\ \bibinfo {author} {\bibfnamefont {M.~A.}\
  \bibnamefont {Eriksson}},\ }\href {\doibase 10.1063/1.2816331} {\bibfield
  {journal} {\bibinfo  {journal} {Applied Physics Letters}\ }\textbf {\bibinfo
  {volume} {91}},\ \bibinfo {pages} {213103} (\bibinfo {year}
  {2007})}\BibitemShut {NoStop}%
\bibitem [{\citenamefont {Field}\ \emph {et~al.}(1993)\citenamefont {Field},
  \citenamefont {Smith}, \citenamefont {Pepper}, \citenamefont {Ritchie},
  \citenamefont {Frost}, \citenamefont {Jones},\ and\ \citenamefont
  {Hasko}}]{Field1993}%
  \BibitemOpen
  \bibfield  {author} {\bibinfo {author} {\bibfnamefont {M.}~\bibnamefont
  {Field}}, \bibinfo {author} {\bibfnamefont {C.~G.}\ \bibnamefont {Smith}},
  \bibinfo {author} {\bibfnamefont {M.}~\bibnamefont {Pepper}}, \bibinfo
  {author} {\bibfnamefont {D.~A.}\ \bibnamefont {Ritchie}}, \bibinfo {author}
  {\bibfnamefont {J.~E.~F.}\ \bibnamefont {Frost}}, \bibinfo {author}
  {\bibfnamefont {G.~A.~C.}\ \bibnamefont {Jones}}, \ and\ \bibinfo {author}
  {\bibfnamefont {D.~G.}\ \bibnamefont {Hasko}},\ }\href {\doibase
  10.1103/PhysRevLett.70.1311} {\bibfield  {journal} {\bibinfo  {journal}
  {Physical Review Letters}\ }\textbf {\bibinfo {volume} {70}},\ \bibinfo
  {pages} {1311} (\bibinfo {year} {1993})}\BibitemShut {NoStop}%
\bibitem [{\citenamefont {Gaudreau}\ \emph {et~al.}(2006)\citenamefont
  {Gaudreau}, \citenamefont {Studenikin}, \citenamefont {Sachrajda},
  \citenamefont {Zawadzki}, \citenamefont {Kam}, \citenamefont {Lapointe},
  \citenamefont {Korkusinski},\ and\ \citenamefont {Hawrylak}}]{Gaudreau2006}%
  \BibitemOpen
  \bibfield  {author} {\bibinfo {author} {\bibfnamefont {L.}~\bibnamefont
  {Gaudreau}}, \bibinfo {author} {\bibfnamefont {S.~A.}\ \bibnamefont
  {Studenikin}}, \bibinfo {author} {\bibfnamefont {A.~S.}\ \bibnamefont
  {Sachrajda}}, \bibinfo {author} {\bibfnamefont {P.}~\bibnamefont {Zawadzki}},
  \bibinfo {author} {\bibfnamefont {A.}~\bibnamefont {Kam}}, \bibinfo {author}
  {\bibfnamefont {J.}~\bibnamefont {Lapointe}}, \bibinfo {author}
  {\bibfnamefont {M.}~\bibnamefont {Korkusinski}}, \ and\ \bibinfo {author}
  {\bibfnamefont {P.}~\bibnamefont {Hawrylak}},\ }\href {\doibase
  10.1103/PhysRevLett.97.036807} {\bibfield  {journal} {\bibinfo  {journal}
  {Physical Review Letters}\ }\textbf {\bibinfo {volume} {97}},\ \bibinfo
  {pages} {036807} (\bibinfo {year} {2006})}\BibitemShut {NoStop}%
\bibitem [{\citenamefont {Petersson}\ \emph {et~al.}(2010)\citenamefont
  {Petersson}, \citenamefont {Petta}, \citenamefont {Lu},\ and\ \citenamefont
  {Gossard}}]{Petersson2010}%
  \BibitemOpen
  \bibfield  {author} {\bibinfo {author} {\bibfnamefont {K.~D.}\ \bibnamefont
  {Petersson}}, \bibinfo {author} {\bibfnamefont {J.~R.}\ \bibnamefont
  {Petta}}, \bibinfo {author} {\bibfnamefont {H.}~\bibnamefont {Lu}}, \ and\
  \bibinfo {author} {\bibfnamefont {A.~C.}\ \bibnamefont {Gossard}},\ }\href
  {\doibase 10.1103/PhysRevLett.105.246804} {\bibfield  {journal} {\bibinfo
  {journal} {Physical Review Letters}\ }\textbf {\bibinfo {volume} {105}},\
  \bibinfo {pages} {246804} (\bibinfo {year} {2010})}\BibitemShut {NoStop}%
\bibitem [{\citenamefont {Korkusinski}\ \emph {et~al.}(2007)\citenamefont
  {Korkusinski}, \citenamefont {Gimenez}, \citenamefont {Hawrylak},
  \citenamefont {Gaudreau}, \citenamefont {Studenikin},\ and\ \citenamefont
  {Sachrajda}}]{Korkusinski2007}%
  \BibitemOpen
  \bibfield  {author} {\bibinfo {author} {\bibfnamefont {M.}~\bibnamefont
  {Korkusinski}}, \bibinfo {author} {\bibfnamefont {I.~P.}\ \bibnamefont
  {Gimenez}}, \bibinfo {author} {\bibfnamefont {P.}~\bibnamefont {Hawrylak}},
  \bibinfo {author} {\bibfnamefont {L.}~\bibnamefont {Gaudreau}}, \bibinfo
  {author} {\bibfnamefont {S.~A.}\ \bibnamefont {Studenikin}}, \ and\ \bibinfo
  {author} {\bibfnamefont {A.~S.}\ \bibnamefont {Sachrajda}},\ }\href {\doibase
  10.1103/PhysRevB.75.115301} {\bibfield  {journal} {\bibinfo  {journal}
  {Physical Review B}\ }\textbf {\bibinfo {volume} {75}},\ \bibinfo {pages}
  {115301} (\bibinfo {year} {2007})}\BibitemShut {NoStop}%
\bibitem [{\citenamefont {Gaudreau}\ \emph {et~al.}(2011)\citenamefont
  {Gaudreau}, \citenamefont {Granger}, \citenamefont {Kam}, \citenamefont
  {Aers}, \citenamefont {Studenikin}, \citenamefont {Zawadzki}, \citenamefont
  {Pioro-Ladri\`{e}re}, \citenamefont {Wasilewski},\ and\ \citenamefont
  {Sachrajda}}]{Gaudreau2011}%
  \BibitemOpen
  \bibfield  {author} {\bibinfo {author} {\bibfnamefont {L.}~\bibnamefont
  {Gaudreau}}, \bibinfo {author} {\bibfnamefont {G.}~\bibnamefont {Granger}},
  \bibinfo {author} {\bibfnamefont {A.}~\bibnamefont {Kam}}, \bibinfo {author}
  {\bibfnamefont {G.~C.}\ \bibnamefont {Aers}}, \bibinfo {author}
  {\bibfnamefont {S.~A.}\ \bibnamefont {Studenikin}}, \bibinfo {author}
  {\bibfnamefont {P.}~\bibnamefont {Zawadzki}}, \bibinfo {author}
  {\bibfnamefont {M.}~\bibnamefont {Pioro-Ladri\`{e}re}}, \bibinfo {author}
  {\bibfnamefont {Z.~R.}\ \bibnamefont {Wasilewski}}, \ and\ \bibinfo {author}
  {\bibfnamefont {A.~S.}\ \bibnamefont {Sachrajda}},\ }\href {\doibase
  10.1038/nphys2149} {\bibfield  {journal} {\bibinfo  {journal} {Nature
  Physics}\ }\textbf {\bibinfo {volume} {8}},\ \bibinfo {pages} {54} (\bibinfo
  {year} {2011})}\BibitemShut {NoStop}%
\bibitem [{\citenamefont {Busl}\ \emph {et~al.}(2013)\citenamefont {Busl},
  \citenamefont {Granger}, \citenamefont {Gaudreau}, \citenamefont
  {S\'{a}nchez}, \citenamefont {Kam}, \citenamefont {Pioro-Ladri\`{e}re},
  \citenamefont {Studenikin}, \citenamefont {Zawadzki}, \citenamefont
  {Wasilewski}, \citenamefont {Sachrajda},\ and\ \citenamefont
  {Platero}}]{Busl2013}%
  \BibitemOpen
  \bibfield  {author} {\bibinfo {author} {\bibfnamefont {M.}~\bibnamefont
  {Busl}}, \bibinfo {author} {\bibfnamefont {G.}~\bibnamefont {Granger}},
  \bibinfo {author} {\bibfnamefont {L.}~\bibnamefont {Gaudreau}}, \bibinfo
  {author} {\bibfnamefont {R.}~\bibnamefont {S\'{a}nchez}}, \bibinfo {author}
  {\bibfnamefont {A.}~\bibnamefont {Kam}}, \bibinfo {author} {\bibfnamefont
  {M.}~\bibnamefont {Pioro-Ladri\`{e}re}}, \bibinfo {author} {\bibfnamefont
  {S.~A.}\ \bibnamefont {Studenikin}}, \bibinfo {author} {\bibfnamefont
  {P.}~\bibnamefont {Zawadzki}}, \bibinfo {author} {\bibfnamefont {Z.~R.}\
  \bibnamefont {Wasilewski}}, \bibinfo {author} {\bibfnamefont {A.~S.}\
  \bibnamefont {Sachrajda}}, \ and\ \bibinfo {author} {\bibfnamefont
  {G.}~\bibnamefont {Platero}},\ }\href {\doibase 10.1038/nnano.2013.7}
  {\bibfield  {journal} {\bibinfo  {journal} {Nature Nanotechnology}\ }\textbf
  {\bibinfo {volume} {8}},\ \bibinfo {pages} {261} (\bibinfo {year}
  {2013})}\BibitemShut {NoStop}%
\bibitem [{\citenamefont {Ribeiro}\ \emph {et~al.}(2013)\citenamefont
  {Ribeiro}, \citenamefont {Burkard}, \citenamefont {Petta}, \citenamefont
  {Lu},\ and\ \citenamefont {Gossard}}]{Ribeiro2013}%
  \BibitemOpen
  \bibfield  {author} {\bibinfo {author} {\bibfnamefont {H.}~\bibnamefont
  {Ribeiro}}, \bibinfo {author} {\bibfnamefont {G.}~\bibnamefont {Burkard}},
  \bibinfo {author} {\bibfnamefont {J.~R.}\ \bibnamefont {Petta}}, \bibinfo
  {author} {\bibfnamefont {H.}~\bibnamefont {Lu}}, \ and\ \bibinfo {author}
  {\bibfnamefont {A.~C.}\ \bibnamefont {Gossard}},\ }\href {\doibase
  10.1103/PhysRevLett.110.086804} {\bibfield  {journal} {\bibinfo  {journal}
  {Physical Review Letters}\ }\textbf {\bibinfo {volume} {110}},\ \bibinfo
  {pages} {086804} (\bibinfo {year} {2013})}\BibitemShut {NoStop}%
\bibitem [{\citenamefont {Lent}\ \emph {et~al.}(1993)\citenamefont {Lent},
  \citenamefont {Tougaw}, \citenamefont {Porod},\ and\ \citenamefont
  {Bernstein}}]{Lent1993}%
  \BibitemOpen
  \bibfield  {author} {\bibinfo {author} {\bibfnamefont {C.~S.}\ \bibnamefont
  {Lent}}, \bibinfo {author} {\bibfnamefont {P.~D.}\ \bibnamefont {Tougaw}},
  \bibinfo {author} {\bibfnamefont {W.}~\bibnamefont {Porod}}, \ and\ \bibinfo
  {author} {\bibfnamefont {G.~H.}\ \bibnamefont {Bernstein}},\ }\href
  {http://iopscience.iop.org/0957-4484/4/1/004} {\bibfield  {journal} {\bibinfo
   {journal} {Nanotechnology}\ }\textbf {\bibinfo {volume} {4}},\ \bibinfo
  {pages} {49} (\bibinfo {year} {1993})}\BibitemShut {NoStop}%
\bibitem [{\citenamefont {Lu}\ \emph {et~al.}(2013)\citenamefont {Lu},
  \citenamefont {Liu},\ and\ \citenamefont {O'Neill}}]{Lu2013}%
  \BibitemOpen
  \bibfield  {author} {\bibinfo {author} {\bibfnamefont {L.}~\bibnamefont
  {Lu}}, \bibinfo {author} {\bibfnamefont {W.}~\bibnamefont {Liu}}, \ and\
  \bibinfo {author} {\bibfnamefont {M.}~\bibnamefont {O'Neill}},\ }\href
  {http://ieeexplore.ieee.org/xpls/abs\_all.jsp?arnumber=6109234} {\bibfield
  {journal} {\bibinfo  {journal} {IEEE Transactions on Computers}\ }\textbf
  {\bibinfo {volume} {62}},\ \bibinfo {pages} {548} (\bibinfo {year}
  {2013})}\BibitemShut {NoStop}%
\bibitem [{\citenamefont {Loss}\ and\ \citenamefont
  {DiVincenzo}(1998)}]{Loss1998}%
  \BibitemOpen
  \bibfield  {author} {\bibinfo {author} {\bibfnamefont {D.}~\bibnamefont
  {Loss}}\ and\ \bibinfo {author} {\bibfnamefont {D.~P.}\ \bibnamefont
  {DiVincenzo}},\ }\href {\doibase 10.1103/PhysRevA.57.120} {\bibfield
  {journal} {\bibinfo  {journal} {Physical Review A}\ }\textbf {\bibinfo
  {volume} {57}},\ \bibinfo {pages} {120} (\bibinfo {year} {1998})}\BibitemShut
  {NoStop}%
\bibitem [{\citenamefont {Fuechsle}\ \emph {et~al.}(2012)\citenamefont
  {Fuechsle}, \citenamefont {Miwa}, \citenamefont {Mahapatra}, \citenamefont
  {Ryu}, \citenamefont {Lee}, \citenamefont {Warschkow}, \citenamefont
  {Hollenberg}, \citenamefont {Klimeck},\ and\ \citenamefont
  {Simmons}}]{Fuechsle2012}%
  \BibitemOpen
  \bibfield  {author} {\bibinfo {author} {\bibfnamefont {M.}~\bibnamefont
  {Fuechsle}}, \bibinfo {author} {\bibfnamefont {J.~A.}\ \bibnamefont {Miwa}},
  \bibinfo {author} {\bibfnamefont {S.}~\bibnamefont {Mahapatra}}, \bibinfo
  {author} {\bibfnamefont {H.}~\bibnamefont {Ryu}}, \bibinfo {author}
  {\bibfnamefont {S.}~\bibnamefont {Lee}}, \bibinfo {author} {\bibfnamefont
  {O.}~\bibnamefont {Warschkow}}, \bibinfo {author} {\bibfnamefont {L.~C.~L.}\
  \bibnamefont {Hollenberg}}, \bibinfo {author} {\bibfnamefont
  {G.}~\bibnamefont {Klimeck}}, \ and\ \bibinfo {author} {\bibfnamefont
  {M.~Y.}\ \bibnamefont {Simmons}},\ }\href {\doibase 10.1038/nnano.2012.21}
  {\bibfield  {journal} {\bibinfo  {journal} {Nature nanotechnology}\ }\textbf
  {\bibinfo {volume} {7}},\ \bibinfo {pages} {1} (\bibinfo {year}
  {2012})}\BibitemShut {NoStop}%
\bibitem [{\citenamefont {Morello}\ \emph {et~al.}(2010)\citenamefont
  {Morello}, \citenamefont {Pla}, \citenamefont {Zwanenburg}, \citenamefont
  {Chan}, \citenamefont {Tan}, \citenamefont {Huebl}, \citenamefont
  {M\"{o}tt\"{o}nen}, \citenamefont {Nugroho}, \citenamefont {Yang},
  \citenamefont {{van Donkelaar}}, \citenamefont {Alves}, \citenamefont
  {Jamieson}, \citenamefont {Escott}, \citenamefont {Hollenberg}, \citenamefont
  {Clark},\ and\ \citenamefont {Dzurak}}]{Morello2010}%
  \BibitemOpen
  \bibfield  {author} {\bibinfo {author} {\bibfnamefont {A.}~\bibnamefont
  {Morello}}, \bibinfo {author} {\bibfnamefont {J.~J.}\ \bibnamefont {Pla}},
  \bibinfo {author} {\bibfnamefont {F.~A.}\ \bibnamefont {Zwanenburg}},
  \bibinfo {author} {\bibfnamefont {K.~W.}\ \bibnamefont {Chan}}, \bibinfo
  {author} {\bibfnamefont {K.~Y.}\ \bibnamefont {Tan}}, \bibinfo {author}
  {\bibfnamefont {H.}~\bibnamefont {Huebl}}, \bibinfo {author} {\bibfnamefont
  {M.}~\bibnamefont {M\"{o}tt\"{o}nen}}, \bibinfo {author} {\bibfnamefont
  {C.~D.}\ \bibnamefont {Nugroho}}, \bibinfo {author} {\bibfnamefont
  {C.}~\bibnamefont {Yang}}, \bibinfo {author} {\bibfnamefont {J.~A.}\
  \bibnamefont {{van Donkelaar}}}, \bibinfo {author} {\bibfnamefont {A.~D.~C.}\
  \bibnamefont {Alves}}, \bibinfo {author} {\bibfnamefont {D.~N.}\ \bibnamefont
  {Jamieson}}, \bibinfo {author} {\bibfnamefont {C.~C.}\ \bibnamefont
  {Escott}}, \bibinfo {author} {\bibfnamefont {L.~C.~L.}\ \bibnamefont
  {Hollenberg}}, \bibinfo {author} {\bibfnamefont {R.~G.}\ \bibnamefont
  {Clark}}, \ and\ \bibinfo {author} {\bibfnamefont {A.~S.}\ \bibnamefont
  {Dzurak}},\ }\href {\doibase 10.1038/nature09392} {\bibfield  {journal}
  {\bibinfo  {journal} {Nature}\ }\textbf {\bibinfo {volume} {467}},\ \bibinfo
  {pages} {687} (\bibinfo {year} {2010})}\BibitemShut {NoStop}%
\bibitem [{\citenamefont {Yin}\ \emph {et~al.}(2013)\citenamefont {Yin},
  \citenamefont {Rancic}, \citenamefont {{de Boo}}, \citenamefont {Stavrias},
  \citenamefont {McCallum}, \citenamefont {Sellars},\ and\ \citenamefont
  {Rogge}}]{Yin2013}%
  \BibitemOpen
  \bibfield  {author} {\bibinfo {author} {\bibfnamefont {C.}~\bibnamefont
  {Yin}}, \bibinfo {author} {\bibfnamefont {M.}~\bibnamefont {Rancic}},
  \bibinfo {author} {\bibfnamefont {G.~G.}\ \bibnamefont {{de Boo}}}, \bibinfo
  {author} {\bibfnamefont {N.}~\bibnamefont {Stavrias}}, \bibinfo {author}
  {\bibfnamefont {J.~C.}\ \bibnamefont {McCallum}}, \bibinfo {author}
  {\bibfnamefont {M.~J.}\ \bibnamefont {Sellars}}, \ and\ \bibinfo {author}
  {\bibfnamefont {S.}~\bibnamefont {Rogge}},\ }\href {\doibase
  10.1038/nature12081} {\bibfield  {journal} {\bibinfo  {journal} {Nature}\
  }\textbf {\bibinfo {volume} {497}},\ \bibinfo {pages} {91} (\bibinfo {year}
  {2013})}\BibitemShut {NoStop}%
\bibitem [{\citenamefont {Smakman}\ \emph {et~al.}(2013)\citenamefont
  {Smakman}, \citenamefont {{van Bree}},\ and\ \citenamefont
  {Koenraad}}]{Smakman2013}%
  \BibitemOpen
  \bibfield  {author} {\bibinfo {author} {\bibfnamefont {E.~P.}\ \bibnamefont
  {Smakman}}, \bibinfo {author} {\bibfnamefont {J.}~\bibnamefont {{van Bree}}},
  \ and\ \bibinfo {author} {\bibfnamefont {P.~M.}\ \bibnamefont {Koenraad}},\
  }\href {\doibase 10.1103/PhysRevB.87.085414} {\bibfield  {journal} {\bibinfo
  {journal} {Physical Review B}\ }\textbf {\bibinfo {volume} {87}},\ \bibinfo
  {pages} {085414} (\bibinfo {year} {2013})}\BibitemShut {NoStop}%
\bibitem [{\citenamefont {Kane}(2005)}]{Kane2005}%
  \BibitemOpen
  \bibfield  {author} {\bibinfo {author} {\bibfnamefont {B.}~\bibnamefont
  {Kane}},\ }\href
  {http://journals.cambridge.org/production/action/cjoGetFulltext?fulltextid=7961932}
  {\bibfield  {journal} {\bibinfo  {journal} {MRS Bulletin}\ }\textbf {\bibinfo
  {volume} {30}},\ \bibinfo {pages} {105} (\bibinfo {year} {2005})}\BibitemShut
  {NoStop}%
\bibitem [{\citenamefont {Haider}\ \emph {et~al.}(2009)\citenamefont {Haider},
  \citenamefont {Pitters}, \citenamefont {DiLabio}, \citenamefont {Livadaru},
  \citenamefont {Mutus},\ and\ \citenamefont {Wolkow}}]{Haider2009}%
  \BibitemOpen
  \bibfield  {author} {\bibinfo {author} {\bibfnamefont {M.~B.}\ \bibnamefont
  {Haider}}, \bibinfo {author} {\bibfnamefont {J.~L.}\ \bibnamefont {Pitters}},
  \bibinfo {author} {\bibfnamefont {G.~A.}\ \bibnamefont {DiLabio}}, \bibinfo
  {author} {\bibfnamefont {L.}~\bibnamefont {Livadaru}}, \bibinfo {author}
  {\bibfnamefont {J.~Y.}\ \bibnamefont {Mutus}}, \ and\ \bibinfo {author}
  {\bibfnamefont {R.~A.}\ \bibnamefont {Wolkow}},\ }\href {\doibase
  10.1103/PhysRevLett.102.046805} {\bibfield  {journal} {\bibinfo  {journal}
  {Physical Review Letters}\ }\textbf {\bibinfo {volume} {102}},\ \bibinfo
  {pages} {046805} (\bibinfo {year} {2009})}\BibitemShut {NoStop}%
\bibitem [{\citenamefont {Pitters}\ \emph {et~al.}(2011)\citenamefont
  {Pitters}, \citenamefont {Livadaru}, \citenamefont {Haider},\ and\
  \citenamefont {Wolkow}}]{Pitters2011a}%
  \BibitemOpen
  \bibfield  {author} {\bibinfo {author} {\bibfnamefont {J.~L.}\ \bibnamefont
  {Pitters}}, \bibinfo {author} {\bibfnamefont {L.}~\bibnamefont {Livadaru}},
  \bibinfo {author} {\bibfnamefont {M.~B.}\ \bibnamefont {Haider}}, \ and\
  \bibinfo {author} {\bibfnamefont {R.~A.}\ \bibnamefont {Wolkow}},\ }\href
  {\doibase 10.1063/1.3514896} {\bibfield  {journal} {\bibinfo  {journal} {The
  Journal of Chemical Physics}\ }\textbf {\bibinfo {volume} {134}},\ \bibinfo
  {pages} {064712} (\bibinfo {year} {2011})}\BibitemShut {NoStop}%
\bibitem [{\citenamefont {Shaterzadeh-Yazdi}\ \emph {et~al.}(2014)\citenamefont
  {Shaterzadeh-Yazdi}, \citenamefont {Livadaru}, \citenamefont {Taucer},
  \citenamefont {Mutus}, \citenamefont {Pitters}, \citenamefont {Wolkow},\ and\
  \citenamefont {Sanders}}]{Yazdi2014}%
  \BibitemOpen
  \bibfield  {author} {\bibinfo {author} {\bibfnamefont {Z.}~\bibnamefont
  {Shaterzadeh-Yazdi}}, \bibinfo {author} {\bibfnamefont {L.}~\bibnamefont
  {Livadaru}}, \bibinfo {author} {\bibfnamefont {M.}~\bibnamefont {Taucer}},
  \bibinfo {author} {\bibfnamefont {J.}~\bibnamefont {Mutus}}, \bibinfo
  {author} {\bibfnamefont {J.~L.}\ \bibnamefont {Pitters}}, \bibinfo {author}
  {\bibfnamefont {R.~A.}\ \bibnamefont {Wolkow}}, \ and\ \bibinfo {author}
  {\bibfnamefont {B.~C.}\ \bibnamefont {Sanders}},\ }\href@noop {} {\bibfield
  {journal} {\bibinfo  {journal} {Physical Review B}\ } (\bibinfo {year}
  {2014})}\BibitemShut {NoStop}%
\bibitem [{\citenamefont {Robles}\ \emph {et~al.}(2012)\citenamefont {Robles},
  \citenamefont {Kepenekian}, \citenamefont {Monturet}, \citenamefont
  {Joachim},\ and\ \citenamefont {Lorente}}]{Robles2012}%
  \BibitemOpen
  \bibfield  {author} {\bibinfo {author} {\bibfnamefont {R.}~\bibnamefont
  {Robles}}, \bibinfo {author} {\bibfnamefont {M.}~\bibnamefont {Kepenekian}},
  \bibinfo {author} {\bibfnamefont {S.}~\bibnamefont {Monturet}}, \bibinfo
  {author} {\bibfnamefont {C.}~\bibnamefont {Joachim}}, \ and\ \bibinfo
  {author} {\bibfnamefont {N.}~\bibnamefont {Lorente}},\ }\href {\doibase
  10.1088/0953-8984/24/44/445004} {\bibfield  {journal} {\bibinfo  {journal}
  {Journal of Physics: Condensed Matter}\ }\textbf {\bibinfo {volume} {24}},\
  \bibinfo {pages} {445004} (\bibinfo {year} {2012})}\BibitemShut {NoStop}%
\bibitem [{\citenamefont {Kawai}\ \emph {et~al.}(2012)\citenamefont {Kawai},
  \citenamefont {Ample}, \citenamefont {Wang}, \citenamefont {Yeo},
  \citenamefont {Saeys},\ and\ \citenamefont {Joachim}}]{Kawai2012}%
  \BibitemOpen
  \bibfield  {author} {\bibinfo {author} {\bibfnamefont {H.}~\bibnamefont
  {Kawai}}, \bibinfo {author} {\bibfnamefont {F.}~\bibnamefont {Ample}},
  \bibinfo {author} {\bibfnamefont {Q.}~\bibnamefont {Wang}}, \bibinfo {author}
  {\bibfnamefont {Y.~K.}\ \bibnamefont {Yeo}}, \bibinfo {author} {\bibfnamefont
  {M.}~\bibnamefont {Saeys}}, \ and\ \bibinfo {author} {\bibfnamefont
  {C.}~\bibnamefont {Joachim}},\ }\href {\doibase
  10.1088/0953-8984/24/9/095011} {\bibfield  {journal} {\bibinfo  {journal}
  {Journal of Physics: Condensed Matter}\ }\textbf {\bibinfo {volume} {24}},\
  \bibinfo {pages} {095011} (\bibinfo {year} {2012})}\BibitemShut {NoStop}%
\bibitem [{\citenamefont {Livadaru}\ \emph {et~al.}(2010)\citenamefont
  {Livadaru}, \citenamefont {Xue}, \citenamefont {Shaterzadeh-Yazdi},
  \citenamefont {DiLabio}, \citenamefont {Mutus}, \citenamefont {Pitters},
  \citenamefont {Sanders},\ and\ \citenamefont {Wolkow}}]{Livadaru2010}%
  \BibitemOpen
  \bibfield  {author} {\bibinfo {author} {\bibfnamefont {L.}~\bibnamefont
  {Livadaru}}, \bibinfo {author} {\bibfnamefont {P.}~\bibnamefont {Xue}},
  \bibinfo {author} {\bibfnamefont {Z.}~\bibnamefont {Shaterzadeh-Yazdi}},
  \bibinfo {author} {\bibfnamefont {G.~A.}\ \bibnamefont {DiLabio}}, \bibinfo
  {author} {\bibfnamefont {J.}~\bibnamefont {Mutus}}, \bibinfo {author}
  {\bibfnamefont {J.~L.}\ \bibnamefont {Pitters}}, \bibinfo {author}
  {\bibfnamefont {B.~C.}\ \bibnamefont {Sanders}}, \ and\ \bibinfo {author}
  {\bibfnamefont {R.~A.}\ \bibnamefont {Wolkow}},\ }\href {\doibase
  10.1088/1367-2630/12/8/083018} {\bibfield  {journal} {\bibinfo  {journal}
  {New Journal of Physics}\ }\textbf {\bibinfo {volume} {12}},\ \bibinfo
  {pages} {083018} (\bibinfo {year} {2010})}\BibitemShut {NoStop}%
\bibitem [{\citenamefont {Bellec}\ \emph {et~al.}(2013)\citenamefont {Bellec},
  \citenamefont {Chaput}, \citenamefont {Dujardin}, \citenamefont {Riedel},
  \citenamefont {Stauffer},\ and\ \citenamefont {Sonnet}}]{Bellec2013}%
  \BibitemOpen
  \bibfield  {author} {\bibinfo {author} {\bibfnamefont {A.}~\bibnamefont
  {Bellec}}, \bibinfo {author} {\bibfnamefont {L.}~\bibnamefont {Chaput}},
  \bibinfo {author} {\bibfnamefont {G.}~\bibnamefont {Dujardin}}, \bibinfo
  {author} {\bibfnamefont {D.}~\bibnamefont {Riedel}}, \bibinfo {author}
  {\bibfnamefont {L.}~\bibnamefont {Stauffer}}, \ and\ \bibinfo {author}
  {\bibfnamefont {P.}~\bibnamefont {Sonnet}},\ }\href {\doibase
  10.1103/PhysRevB.88.241406} {\bibfield  {journal} {\bibinfo  {journal}
  {Physical Review B}\ }\textbf {\bibinfo {volume} {88}},\ \bibinfo {pages}
  {241406} (\bibinfo {year} {2013})}\BibitemShut {NoStop}%
\bibitem [{\citenamefont {Schofield}\ \emph {et~al.}(2013)\citenamefont
  {Schofield}, \citenamefont {Studer}, \citenamefont {Hirjibehedin},
  \citenamefont {Curson}, \citenamefont {Aeppli},\ and\ \citenamefont
  {Bowler}}]{Schofield2013}%
  \BibitemOpen
  \bibfield  {author} {\bibinfo {author} {\bibfnamefont {S.~R.}\ \bibnamefont
  {Schofield}}, \bibinfo {author} {\bibfnamefont {P.}~\bibnamefont {Studer}},
  \bibinfo {author} {\bibfnamefont {C.~F.}\ \bibnamefont {Hirjibehedin}},
  \bibinfo {author} {\bibfnamefont {N.~J.}\ \bibnamefont {Curson}}, \bibinfo
  {author} {\bibfnamefont {G.}~\bibnamefont {Aeppli}}, \ and\ \bibinfo {author}
  {\bibfnamefont {D.~R.}\ \bibnamefont {Bowler}},\ }\href {\doibase
  10.1038/ncomms2679} {\bibfield  {journal} {\bibinfo  {journal} {Nature
  Communications}\ }\textbf {\bibinfo {volume} {4}},\ \bibinfo {pages} {1649}
  (\bibinfo {year} {2013})}\BibitemShut {NoStop}%
\bibitem [{\citenamefont {Goh}\ \emph {et~al.}(2011)\citenamefont {Goh},
  \citenamefont {Chen}, \citenamefont {Xu}, \citenamefont {Ballard},
  \citenamefont {Randall},\ and\ \citenamefont {{Von Ehr}}}]{Goh2011}%
  \BibitemOpen
  \bibfield  {author} {\bibinfo {author} {\bibfnamefont {K.~E.~J.}\
  \bibnamefont {Goh}}, \bibinfo {author} {\bibfnamefont {S.}~\bibnamefont
  {Chen}}, \bibinfo {author} {\bibfnamefont {H.}~\bibnamefont {Xu}}, \bibinfo
  {author} {\bibfnamefont {J.}~\bibnamefont {Ballard}}, \bibinfo {author}
  {\bibfnamefont {J.~N.}\ \bibnamefont {Randall}}, \ and\ \bibinfo {author}
  {\bibfnamefont {J.~R.}\ \bibnamefont {{Von Ehr}}},\ }\href {\doibase
  10.1063/1.3582241} {\bibfield  {journal} {\bibinfo  {journal} {Applied
  Physics Letters}\ }\textbf {\bibinfo {volume} {98}},\ \bibinfo {pages}
  {163102} (\bibinfo {year} {2011})}\BibitemShut {NoStop}%
\bibitem [{\citenamefont {Chen}\ \emph {et~al.}(2012)\citenamefont {Chen},
  \citenamefont {Xu}, \citenamefont {Goh}, \citenamefont {Liu},\ and\
  \citenamefont {Randall}}]{Chen2012}%
  \BibitemOpen
  \bibfield  {author} {\bibinfo {author} {\bibfnamefont {S.}~\bibnamefont
  {Chen}}, \bibinfo {author} {\bibfnamefont {H.}~\bibnamefont {Xu}}, \bibinfo
  {author} {\bibfnamefont {K.~E.~J.}\ \bibnamefont {Goh}}, \bibinfo {author}
  {\bibfnamefont {L.}~\bibnamefont {Liu}}, \ and\ \bibinfo {author}
  {\bibfnamefont {J.~N.}\ \bibnamefont {Randall}},\ }\href {\doibase
  10.1088/0957-4484/23/27/275301} {\bibfield  {journal} {\bibinfo  {journal}
  {Nanotechnology}\ }\textbf {\bibinfo {volume} {23}},\ \bibinfo {pages}
  {275301} (\bibinfo {year} {2012})}\BibitemShut {NoStop}%
\bibitem [{\citenamefont {Kolmer}\ \emph {et~al.}(2013)\citenamefont {Kolmer},
  \citenamefont {Godlewski}, \citenamefont {Zuzak}, \citenamefont {Wojtaszek},
  \citenamefont {Rauer}, \citenamefont {Thuaire}, \citenamefont {Hartmann},
  \citenamefont {Moriceau}, \citenamefont {Joachim},\ and\ \citenamefont
  {Szymonski}}]{Kolmer2013}%
  \BibitemOpen
  \bibfield  {author} {\bibinfo {author} {\bibfnamefont {M.}~\bibnamefont
  {Kolmer}}, \bibinfo {author} {\bibfnamefont {S.}~\bibnamefont {Godlewski}},
  \bibinfo {author} {\bibfnamefont {R.}~\bibnamefont {Zuzak}}, \bibinfo
  {author} {\bibfnamefont {M.}~\bibnamefont {Wojtaszek}}, \bibinfo {author}
  {\bibfnamefont {C.}~\bibnamefont {Rauer}}, \bibinfo {author} {\bibfnamefont
  {A.}~\bibnamefont {Thuaire}}, \bibinfo {author} {\bibfnamefont {J.-M.}\
  \bibnamefont {Hartmann}}, \bibinfo {author} {\bibfnamefont {H.}~\bibnamefont
  {Moriceau}}, \bibinfo {author} {\bibfnamefont {C.}~\bibnamefont {Joachim}}, \
  and\ \bibinfo {author} {\bibfnamefont {M.}~\bibnamefont {Szymonski}},\ }\href
  {\doibase 10.1016/j.apsusc.2013.09.124} {\bibfield  {journal} {\bibinfo
  {journal} {Applied Surface Science}\ }\textbf {\bibinfo {volume} {288}},\
  \bibinfo {pages} {83} (\bibinfo {year} {2013})}\BibitemShut {NoStop}%
\bibitem [{\citenamefont {Livadaru}\ \emph {et~al.}(2011)\citenamefont
  {Livadaru}, \citenamefont {Pitters}, \citenamefont {Taucer},\ and\
  \citenamefont {Wolkow}}]{Livadaru2011}%
  \BibitemOpen
  \bibfield  {author} {\bibinfo {author} {\bibfnamefont {L.}~\bibnamefont
  {Livadaru}}, \bibinfo {author} {\bibfnamefont {J.~L.}\ \bibnamefont
  {Pitters}}, \bibinfo {author} {\bibfnamefont {M.}~\bibnamefont {Taucer}}, \
  and\ \bibinfo {author} {\bibfnamefont {R.~A.}\ \bibnamefont {Wolkow}},\
  }\href {\doibase 10.1103/PhysRevB.84.205416} {\bibfield  {journal} {\bibinfo
  {journal} {Physical Review B}\ }\textbf {\bibinfo {volume} {84}},\ \bibinfo
  {pages} {205416} (\bibinfo {year} {2011})}\BibitemShut {NoStop}%
\bibitem [{\citenamefont {Rezeq}\ \emph {et~al.}(2006)\citenamefont {Rezeq},
  \citenamefont {Pitters},\ and\ \citenamefont {Wolkow}}]{Rezeq2006}%
  \BibitemOpen
  \bibfield  {author} {\bibinfo {author} {\bibfnamefont {M.}~\bibnamefont
  {Rezeq}}, \bibinfo {author} {\bibfnamefont {J.~L.}\ \bibnamefont {Pitters}},
  \ and\ \bibinfo {author} {\bibfnamefont {R.~A.}\ \bibnamefont {Wolkow}},\
  }\href {\doibase 10.1063/1.2198536} {\bibfield  {journal} {\bibinfo
  {journal} {The Journal of Chemical Physics}\ }\textbf {\bibinfo {volume}
  {124}},\ \bibinfo {pages} {204716} (\bibinfo {year} {2006})}\BibitemShut
  {NoStop}%
\bibitem [{\citenamefont {Boland}(1993)}]{Boland1993}%
  \BibitemOpen
  \bibfield  {author} {\bibinfo {author} {\bibfnamefont {J.~J.}\ \bibnamefont
  {Boland}},\ }\href {\doibase 10.1080/00018739300101474} {\bibfield  {journal}
  {\bibinfo  {journal} {Advances in Physics}\ }\textbf {\bibinfo {volume}
  {42}},\ \bibinfo {pages} {129} (\bibinfo {year} {1993})}\BibitemShut
  {NoStop}%
\bibitem [{\citenamefont {Pitters}\ \emph {et~al.}(2012)\citenamefont
  {Pitters}, \citenamefont {Piva},\ and\ \citenamefont {Wolkow}}]{Pitters2012}%
  \BibitemOpen
  \bibfield  {author} {\bibinfo {author} {\bibfnamefont {J.~L.}\ \bibnamefont
  {Pitters}}, \bibinfo {author} {\bibfnamefont {P.~G.}\ \bibnamefont {Piva}}, \
  and\ \bibinfo {author} {\bibfnamefont {R.~A.}\ \bibnamefont {Wolkow}},\
  }\href@noop {} {\bibfield  {journal} {\bibinfo  {journal} {Journal of Vacuum
  Science \& Technology B}\ }\textbf {\bibinfo {volume} {30}},\ \bibinfo
  {pages} {1} (\bibinfo {year} {2012})}\BibitemShut {NoStop}%
\bibitem [{\citenamefont {Hoffmann}\ and\ \citenamefont
  {Woodside}()}]{Hoffmann2013}%
  \BibitemOpen
  \bibfield  {author} {\bibinfo {author} {\bibfnamefont {A.}~\bibnamefont
  {Hoffmann}}\ and\ \bibinfo {author} {\bibfnamefont {M.~T.}\ \bibnamefont
  {Woodside}},\ }\href@noop {} {}\bibinfo {howpublished} {personal
  communication}\BibitemShut {NoStop}%
\bibitem [{\citenamefont {Hoffmann}\ \emph {et~al.}(2011)\citenamefont
  {Hoffmann}, \citenamefont {Nettels}, \citenamefont {Clark}, \citenamefont
  {Borgia}, \citenamefont {Radford}, \citenamefont {Clarke},\ and\
  \citenamefont {Schuler}}]{Hoffmann2011}%
  \BibitemOpen
  \bibfield  {author} {\bibinfo {author} {\bibfnamefont {A.}~\bibnamefont
  {Hoffmann}}, \bibinfo {author} {\bibfnamefont {D.}~\bibnamefont {Nettels}},
  \bibinfo {author} {\bibfnamefont {J.}~\bibnamefont {Clark}}, \bibinfo
  {author} {\bibfnamefont {A.}~\bibnamefont {Borgia}}, \bibinfo {author}
  {\bibfnamefont {S.~E.}\ \bibnamefont {Radford}}, \bibinfo {author}
  {\bibfnamefont {J.}~\bibnamefont {Clarke}}, \ and\ \bibinfo {author}
  {\bibfnamefont {B.}~\bibnamefont {Schuler}},\ }\href {\doibase
  10.1039/c0cp01911a} {\bibfield  {journal} {\bibinfo  {journal} {Physical
  Chemistry Chemical Physics}\ }\textbf {\bibinfo {volume} {13}},\ \bibinfo
  {pages} {1857} (\bibinfo {year} {2011})}\BibitemShut {NoStop}%
\bibitem [{\citenamefont {Hoffmann}\ and\ \citenamefont
  {Woodside}(2011)}]{Hoffmann2011a}%
  \BibitemOpen
  \bibfield  {author} {\bibinfo {author} {\bibfnamefont {A.}~\bibnamefont
  {Hoffmann}}\ and\ \bibinfo {author} {\bibfnamefont {M.~T.}\ \bibnamefont
  {Woodside}},\ }\href {\doibase 10.1002/anie.201104033} {\bibfield  {journal}
  {\bibinfo  {journal} {Angewandte Chemie (International ed. in English)}\
  }\textbf {\bibinfo {volume} {50}},\ \bibinfo {pages} {12643} (\bibinfo {year}
  {2011})}\BibitemShut {NoStop}%
\end{thebibliography}

%

\end{document}